\documentclass[%
preprint,
 amsmath,amssymb,
 aps,
pre
]{revtex4-2}

\usepackage{graphicx}
\usepackage{dcolumn}
\usepackage{bm}

\newcommand{\kB}{k_{\rm B}}
\newcommand{\be}{\begin{equation}}
\newcommand{\ee}{\end{equation}}
\newcommand{\ba}{\begin{eqnarray}}
\newcommand{\ea}{\end{eqnarray}}

\begin{document}

\preprint{
}

\title{ Second moments of work and heat 
for a single particle stochastic heat engine
in a breathing harmonic potential
}

\author{Shinji IIDA}
\email{iida@rins.ryukoku.ac.jp}
\author{Koki NEGORO}
\author{Kanta YAMADA}
\affiliation{%
Applied Mathematics and Informatics Course, Faculty of Advanced Science and Technology
\\
Ryukoku University, Otsu, Shiga 520-2194, 
Japan 
}%

\date{\today}

\begin{abstract}%
We consider a simple model of a stochastic heat engine, 
which consists of a single Brownian particle moving 
in a one-dimensional periodically breathing  harmonic potential. 
Overdamped limit is assumed.
Expressions of second moments (variances and covariances ) of heat and work 
are obtained in the form of integrals, 
whose integrands contain 
functions satisfying certain differential equations. 
The results in the quasi-static limit are simple functions of 
temperatures of hot and cold thermal baths.
The coefficient of variation of the work is suggested to 
give an approximate probability for the work to 
exceeds a given threshold.
During derivation, 
we get the expression of the cumulant-generating function.
\end{abstract}


\maketitle

\section{\label{sec:introduction}Introduction}

With the progress of microfabrication and measurement technology, 
there has arisen much interest in the properties of artificial or 
biological molecular machines.
A number of research has been made theoretically and 
experimentally ( see Refs.~\cite{Seifert2012,Holubec2021} 
and references therein).
Especially a single particle stochastic heat engine, 
that was realized experimentally by an optically trapped colloidal 
particle (see, e.g., Ref.~\cite{Blickle2012}), has 
invoked a lot of work. 

For example, 
the efficiency of a heat engine at maximal power 
was discussed in Refs.~\cite{Schmiedl2007,Schmiedl2008}.
There the efficiency of a stochastic heat engine was defined 
as the ratio of the average of work to the average of heat.
Since fluctuations often dominate averages 
in such  small-scale systems~\cite{Rana2014,Pal2016},
it is important to understand what kind of role these fluctuations 
play in the performance of stochastic heat engines.
Toward this goal, we present in this paper 
second moments (variances and covariances) 
of heat and work 
for a simple model of a stochastic heat engine 
used in Ref.~\cite{Schmiedl2008}.

Statistical properties of a stochastic heat 
engine have been analyzed in Refs.~\cite{Rana2014,Pal2016}, 
where probability distributions of heat and work were 
obtained by numerical simulations.
Further a distribution of the efficinecy  
was calculated and compared with the analytical expression 
obtained in Ref.~\cite{Polettini2015}.
The variance of work was analytically calculated 
with using of Green's function in Ref.~\cite{Holubec2018}.
We give the analytical results of 
all the second moments of heat and work 
which still seem absent in literature.

Refs.~\cite{Speck2011,Pal2013,Pal2014,Xu2018} have discussed 
statistical poperties of work done on a Brownian particle 
immersed in a single heat bath.
Here in this paper, we consider a heat engine
which contacts with two heat baths at different temperatures.

This paper is organized as follows:
Sect.~\ref{sec:model} describes the model.
The joint probability density is given in Sect.~\ref{sec:probability}.
There, heat and work are defined.
Sect.~\ref{sec:logM} gives the expression of a cumulant-generating function.
Differentiating this function, we calculate second moments 
in Sect.~\ref{sec:2nd}.
Sect.~\ref{sec:discussion} gives figures which show 
how these quantities approach their quasi-static limits.
Sect.~\ref{sec:summary}  summarizes  our results and 
comments on further study.

\section{\label{sec:model}Model}

We use a model of a stochastic heat engine introduced 
in Ref.~\cite{Schmiedl2008}:
a Brownian particle moving along $x$-axis is trapped 
in a harmonic potential,
\be\label{eq:U}
 U(x,t) = \frac{\lambda(t)}{2} x^2
\,.
\ee
The stiffness of the potential $\lambda (t )$ varies 
periodically in time.
The equation of motion for the particle is given by
\be\label{eq:EOM-1}
m \ddot{x}(t) = -\cfrac{\partial U}{\partial x}(x,t)
-\gamma \dot{x}(t) + \xi (t)
\ee
where $m$ represents the mass and $\gamma$ is the friction coefficient.
Here and hereafter a dot means time-derivative.
The noise $\xi (t)$ is Gaussian with zero mean and is delta correlated:
\be\label{eq:xi}
 \langle \xi(t) \rangle = 0\,,\quad
 \langle \xi(t)\xi(t') \rangle = 2\gamma \kB T \delta(t-t')
\ee
where $T$ is the temperature of the thermal bath and
the ensemble average is denoted by $\langle \phantom{x} \rangle$.
For the sake of simplicity, 
we hereafter set the Boltzmann constant $\kB$, to be unity.

We further assume the overdamped limit and neglect 
the inertia term $m \ddot{x}$ in Eq.~(\ref{eq:EOM-1}) 
and start with the following overdamped 
Langevin equation:
\be\label{eq:EOM-2}
 \gamma \dot{x}(t) =  - \lambda (t) x + \xi(t)
\,.
\ee
Although $\gamma$ can be eliminated from the equation 
by rescaling the time as $t = \gamma \tau$, 
we keep this symbol 
in order to consider the quasi-static limit
which is equivalent to taking the limit $\gamma\to 0$.

The cycle consists of two isothermal steps and
two adiabatic steps:

\begin{enumerate}
\item
Isothermal transition at temperature $T_h$ during time
interval $0 < t < t_1$ .
$\lambda(t)$ decreases and the system expands.

\item
Adiabatic transition instantaneously from temperature $T_h$
 to temperature $T_c$ at time $t = t_1$. 
$\lambda(t)$ decreases discontinuously.

\item
Isothermal transition at temperature $T_c$ ($<T_h$)
during time interval $t_1 < t < t_1 + t_3$.
$\lambda(t)$ increases and the system is compressed.

\item
Adiabatic transition from temperature $T_c$ to 
temperature $T_h$ 
at time $t = t_1 + t_3$.
$\lambda(t)$ increases discontinuously.
\end{enumerate}

The system repeats these four steps.

In Ref.~\cite{Schmiedl2008}, 
the probability distribution for the particle to be found 
at position $x$ at time $t$, 
$p(x, t)$ is shown to remain Gaussian with 0 mean and 
the variance $w(t) = \langle X^2(t)\rangle$
for all times $t$ if it is so initially:
\be\label{eq:pxt}
p(x, t) = \frac{1}{\sqrt{2\pi w(t)}}
\exp\left(- \frac{x^2}{2 w(t)}\right)
\,,
\ee
where $X(t)$ denotes a stochastic variable representing  
the position of the particle at the time $t$.

The time evolution of $w(t)$ is given by
\be\label{eq:dwdt}
  \dot{w}(t) = \frac{2}{\gamma}\left( T(t) - \lambda(t)~w(t)\right)
\,.
\ee

For the numerical calculation in Sect.~\ref{sec:discussion}, 
we use the protocol for the variance $w(t)$:
\ba\label{eq:protocol}
w(t) &=& \left\{\begin{array}{ccc}
 w_a \left(
1 + \left(\sqrt{\frac{w_b}{w_a}}-1 \right)~
\frac{t}{t_1} \right)^2& ; & 0<t<t_1
\\
w_b\left(1 + \left(\sqrt{\frac{w_a}{w_b}}-1 \right)~
\frac{t-t_1}{t_3} \right)^2& ; & t_1<t<t_1+t_3
\end{array}
\right.
\,.
\ea
If the time-interval spent in the isothermal transitions, $t_1$ and $t_3$, 
are set to
\be\label{eq:t1t3}
 t_1^{*} = t_3^{*} = \frac{8\gamma 
\left(\sqrt{w_b}- \sqrt{w_a} \right)^2}{(T_h-T_c)\log(w_b/w_a)}
\,,
\ee
the system gives a maximal power output 
for given $T_h$, $T_c$, $w_a$, and $w_b$.
See Fig.2. in Ref.~\cite{Schmiedl2008}
for the profile of $w(t)$ and $\lambda(t)$. 

In order to analyze the quasi-static limit, 
we use
\be\label{eq:s}
 t_1 = s~t_1^{*}\,,\quad t_2 = s~t_2^{*}
\,,
\ee
and vary the value of $s$.

\section{\label{sec:probability}Probability density}

In order to make the presentation simple,
we first divide the time-interval
into $n$ segments whose length is $\Delta t$,  
and consider the limit $\Delta t \to 0$ ($n \to \infty$) in the final stage.

$X_j$ denotes a stochastic variable representing  the position of the particle 
at $t_j = j \Delta t$.
Since $\lambda(t)$ is discontinuous at $t=0$, $t=t_1$, and $t=t_1+t_3$, 
we use the notation
\be\label{eq:t_n}
t_0 = +0\,,\quad
t_{n_1} = t_1 + 0\,,\quad
t_n = t_1 + t_3 +0
\ee 
in order to avoid ambiguity.
We also use the following notations:
\be\label{eq:lambda_j}
 \lambda_j = \lambda(t_j)\,,\quad 
T_j = T(t_j)\,,\quad w_j = w(t_j)
\ee
and
\be\label{eq:wawb}
 w_a = w_0 = w(0)\,,\quad w_b = w_{n_1} = w(t_1)
\,.
\ee 
Note $w(t)$ is a continuous function.

The stochastic differential equation corresponding to 
Eq.~(\ref{eq:EOM-2}) becomes
\be\label{eq:SDE}
\gamma \left( X_{j+1} - X_j \right) = 
- \lambda_j X_j ~\Delta t +\sqrt{2\gamma T_j} \Delta B_j
\ee
where $\Delta B_j$ obeys a Gaussian distribution:
\be\label{eq:pDBt}
 p(\Delta B_t) = \frac{1}{\sqrt{2\pi\Delta t}} 
\exp\left( -\frac{(\Delta B_t)^2}{2\Delta t} \right)
\,.
\ee

The conditional probability density for
$X_{j+1}= x_{j+1}$ with the condition $X_{j}= x_{j}$
is 
\ba\label{eq:conditionalProbability}
p(x_{j+1}\vert x_j) &=& 
\frac{1}{\sqrt{2\pi \Delta t}} 
\exp\left( -\frac{(\Delta B_j)^2}{2\Delta t} \right) 
\sqrt{\frac{\gamma}{2 T_j}}
\nonumber\\
&=&  \sqrt{\frac{\gamma}{4\pi T_j \Delta t}}
\exp\left( -\frac{\left(\gamma (x_{j+1} - x_j)  
+ \lambda_j~x_j~\Delta t \right)^2}
{4\Delta t \gamma T_j} \right) 
\,.
\ea
The joint probability density for
$\{ X_j = x_j ; j=0, \cdots , n\}$ is
\ba\label{eq:jointProbability}
&&p(x_0,x_1,\cdots , x_n) = 
p(x_n|x_{n-1}) p(x_{n-1}|x_{n-2}) \cdots p(x_1|x_0) p(x_0,t_0)
\nonumber\\
&& = 
\Pi_{j=0}^{n-1}
\sqrt{\frac{\gamma}{4\pi T_j \Delta t}}
\exp\left( -\frac{\left(\gamma (x_{j+1} - x_{j})  
+ \lambda_j~x_j~\Delta t \right)^2}{4\Delta t \gamma T_j} \right)
~\frac{\exp\left(- \frac{x_0^2}{2 w_0}\right)}{\sqrt{2\pi w_0}} 
\,.
\ea

Thermodynamic quantities such as work, heat, and internal energy 
can be defined on a single stochastic trajectory
 according to the formalism of stochastic thermodynamics 
\cite{Sekimoto1997, Sekimoto1998, Sekimoto2010}.
We define 
the heat $Q_h$ ($Q_c$) uptake from the hotter ( cooler ) 
heat bath at temperature $T_h$ ($T_c$) as follows:
\ba
 Q_h &=&  
\sum_{j=0}^{n_1 - 1} 
\frac{\lambda_j}{2} \left( X_{j+1}^2 - X_{j}^2 \right) 
\,,
\label{eq:Qh}
\\
 Q_c &=&  
 \sum_{j=n_1}^{n-1} 
\frac{\lambda_j}{2} \left( X_{j+1}^2 - X_{j}^2 \right) 
\,.
\label{eq:Qc}
\ea
The change in internal energy during one cycle is
\be\label{eq:dU}
\Delta U = \frac{\lambda_{n}}{2} X_{n}^2 - \frac{\lambda_{0}}{2} X_{0}^2 
\,.
\ee
The work done by the particle during one cycle is
\be\label{eq:W}
W = - \frac{1}{2} 
\sum_{j=0}^{n-1} X_{j+1}^2 \left( \lambda_{j+1} - \lambda_j \right)
\,.
\ee
The energy conservation holds for an individual stochastic trajectory:
\be\label{eq:1st}
\Delta U = Q_h + Q_c - W
\,.
\ee

\section{\label{sec:logM}Moment- and Cumulant-Generating Function}

With the use of Eqs.~(\ref{eq:jointProbability})$\sim$
(\ref{eq:dU}), 
the moment-generating function of $Q_h$, $Q_c$, and $\Delta U$ is defined as
\ba\label{eq:M}
&&M(S_h,S_c,S_u) = 
\left\langle e^{S_h Q_h + S_c Q_c+S_u\Delta U}\right\rangle
= 
\Pi_{i=0}^{n} \int_{-\infty}^{\infty} dx_i~
\sqrt{\frac{\gamma}{4\pi T_i \Delta t}}
\nonumber\\
&& \quad\quad\quad
\exp\left(-\frac{\left(\gamma (x_{i+1} - x_{i})  
+ \lambda_i~x_i~\Delta t \right)^2}{4\Delta t \gamma T_i}\right)~
\frac{e^{- \frac{x_0^2}{2 w_0}}}{\sqrt{2\pi w_0}} 
~e^{S_h Q_h + S_c Q_c+S_u\Delta U}
\,.
\ea
In order to calculate $M(S_h,S_c,S_u)$, we first define,
 for $j=0,\cdots,n-1$,  
\ba
&&h_{j+1}(x_{j+1}) 
=
 \Pi_{i=0}^{j} \int_{-\infty}^{\infty} dx_i~
\sqrt{\frac{\gamma}{4\pi T_i \Delta t}}
\exp\left( -\frac{\left(\gamma (x_{i+1} - x_{i})  
+ \lambda_i~x_i~\Delta t \right)^2}{4\Delta t \gamma T_i} \right)
\nonumber\\
&&\quad\quad
\frac{1}{\sqrt{2\pi w_0}} \exp\left(- \frac{x_0^2}{2 w_0}
-\frac{\lambda_0}{2}x_0^2 S_u\right)
\exp\left( 
\sum_{k=0}^{j} \frac{\lambda_k}{2}\left(x_{k+1}^2 - x_{k}^2 \right) 
g(t_k)
\right)
\,,
\label{eq:hj+1}
\ea
where
\ba
g(t) &=& \left\{\begin{array}{ccc}
 S_h & ; & 0 < t < t_1
\\
S_c & ; & t_1 < t < t_1+t_3
\end{array}
\right.
\label{eq:gt}
\ea
with 
\be\label{eq:h0}
h_0(x_0) = \frac{1}{\sqrt{2\pi w_0}} \exp\left(- \frac{x_0^2}{2 w_0}
-\frac{\lambda_0}{2}x_0^2 S_u\right)
\,.
\ee

By induction, we can see that $h_{j}(x_{j})$ has the following form:
\be\label{eq:hj}
h_{j}(x_{j})= \frac{1}{\sqrt{2\pi a_{j}}} 
\exp\left(- \frac{x_j^2}{2 b_j}\right)
\,.
\ee
Eq.~(\ref{eq:h0}) gives
\be
 a_0 = w_0 = w_a\,,\quad b_0 = w_a( 1 + S_u \lambda_0 w_a)^{-1}
\,.
\ee

Then the moment-generating function is expressed as
\ba
M(S_h,S_c,S_u) &=& \int_{-\infty}^{\infty}dx_n h_{n}(x_n)\exp\left(\frac{\lambda_n}{2} x_n^2 S_u\right) 
\nonumber\\
 &=& \frac{1}{\sqrt{2\pi a_n}}
\int_{-\infty}^{\infty}dx_n \exp\left(- \frac{1}{2 b_n} x_{n}^2 \right)
\exp\left(\frac{\lambda_n}{2} x_n^2 S_u\right)
\nonumber\\
&=& \sqrt{\frac{b_n \left( 1 - b_n \lambda_n  S_u \right)^{-1}}{a_n}}
\,.
\label{eq:M2}
\ea
From the following equation,
\ba
&&h_{j+1}(x_{j+1}) = \sqrt{\frac{\gamma}{4\pi T_j \Delta t}} 
\int_{-\infty}^{\infty}dx_j h_{j}(x_{j}) 
\nonumber\\
&&\quad\quad\quad
\exp\left( -\frac{\left(\gamma (x_{j+1} - x_j ) + \lambda_{j}~x_j~\Delta t 
\right)^2}{4\Delta t \gamma T_{j}} 
+ \frac{\lambda_j}{2}\left(x_{j+1}^2 - x_{j}^2 
\right) g(t_j)
\right)
\nonumber\\
&&= \sqrt{\frac{\gamma}{4\pi T_j \Delta t}}
\frac{1}{\sqrt{2\pi a_{j}}}
\int_{-\infty}^{\infty} dx_j
\exp\left(- \frac{x_j^2}{2 b_j}\right)
\nonumber\\
&&\quad\quad\quad
\exp\left( -\frac{\left(\gamma (x_{j+1} - x_j ) + \lambda_{j}~x_j~
\Delta t \right)^2}{4\Delta t \gamma T_{j}} 
+ \frac{\lambda_j}{2}\left(x_{j+1}^2 - x_{j}^2 
\right) g(t_j)
\right)
\label{eq:M-1}
\,,
\ea
we can obtain recurrence relations for $a_j$ and $b_j$: 
\be
 a_{j+1} = \frac{2 \Delta t T_j}{\gamma}
\left(\frac{1}{b_j} + \lambda_j g(t_j) 
+ \frac{\left( \gamma - \lambda_j \Delta t\right)^2}{2 \Delta t \gamma T_j}
\right) a_j
\,,
\ee
\be
\frac{1}{b_{j+1}} = - \lambda_j g(t_j) + \frac{\gamma}{2\Delta t T_j}
 - \frac{\left( \gamma - \lambda_j \Delta t\right)^2}{4 (\Delta t T_j)^2}
\left( \frac{1}{b_j} + \lambda_j g(t_j) + 
\frac{\left( \gamma - \lambda_j \Delta t\right)^2}{2 \Delta t \gamma T_j}
\right)^{-1}
\,.
\ee

Considering the limit $\Delta t \to 0$, 
we can get equations 
\be\label{eq:dbdt}
\dot{a} = \frac{2 a}{\gamma}
\frac{( 1 + g \lambda b) T - \lambda b}{b}
\,,\quad
\dot{b} = \frac{2}{\gamma}( 1+ g\lambda b ) 
\left( ( 1 + g \lambda b) T - \lambda b \right)
\ee
with initial conditions
\ba
 a(0)=w_a\,,\quad  b(0) = w_a( 1 + S_u \lambda(+0) w_a)^{-1}
\,.
\label{eq:InitialConditionOfab}
\ea
The moment-generating function, Eq.~(\ref{eq:M2}), becomes
\ba
M(S_h,S_c,S_u)
&=&\left.\sqrt{\frac{b(t) ( 1 - b(t) \lambda(t) S_u)^{-1}}{a(t)}}
\right\vert_{t=t_1+t_3+0}
\,.
\ea
We note that if we set 
${\bm S} = (S_h,S_c,S_u) = {\bm 0}$,
both $a(t)$ and $b(t)$ become $w(t)$, 
and $h_j(x_j)$ reduces to the probability distribution of 
the particle,
 $p(x_j,t_j)$.

The cumulant generating function is
\ba
\log M(S_h,S_c,S_u) &=& 
\log(\hat{M}(t_1+t_3+0))
\nonumber\\ 
&& - \frac{1}{2}\log\left( 1 - b(t_1+t_3+0) \lambda(t_1+t_3+0) S_u \right)
\label{eq:logM-1}
\ea
where
\be
\hat{M}(t) = \sqrt{\frac{b(t)}{a(t)}}
\,.
\ee
Eqs.~(\ref{eq:dbdt}) give
\be\label{eq:dlogMdt}
 \frac{d}{dt} \log\hat{M} = 
 \frac{1}{2}\left(\frac{\frac{db}{dt}}{b} - \frac{\frac{da}{dt}}{a}\right)
= \frac{1}{\gamma}g\lambda \left( ( 1 + g \lambda b) T - \lambda b \right)
\,,
\ee
which leads to 
\ba
&&\log \hat{M}(t_1+t_3+0) = 
\log\frac{\hat{M}(t_1+t_3+0)}{\hat{M}(+0)} + \log\hat{M}(+0)
\nonumber\\
&& \quad =
\frac{1}{\gamma} \int_{0}^{t_1+t_3} g(t') \lambda(t')
\left(
T(t') + \lambda(t') b(t')\Big( g(t')  T(t') -1 \Big) \right) dt'
\nonumber\\
&& \quad\quad
 - \frac{1}{2}\log\left( 1 + S_u \lambda(+0) w_a \right)
\,.
\label{eq:logM-2}
\ea

Eqs.~(\ref{eq:logM-1}) and (\ref{eq:logM-2}) give the expression of the 
cumulant-generating function:
\ba
\log M(S_h,S_c,S_u) &=& 
-\frac{1}{2}\log( 1 + S_u \lambda(+0) w_a)
 - \frac{1}{2}\log( 1 - b(t_1+t_3+0) \lambda(t_1+t_3+0) S_u)
\nonumber\\
&&\quad
+ \frac{S_h T_h }{\gamma} \int_{0}^{t_1} \lambda(t') dt'
+ 
\frac{S_h^2  T_h -S_h}{\gamma} \int_{0}^{t_1} \lambda^2(t') b(t') dt'
\nonumber\\
&&\quad
+ \frac{S_c T_c}{\gamma} \int_{t_1}^{t_1+t_3} \lambda(t') dt'
+ 
\frac{S_c^2  T_{c} -S_c}{\gamma} \int_{t_1}^{t_1+t_3} \lambda^2(t') b(t') dt'
\,,
\label{eq:logM-3}
\ea
which, together with Eqs.~(\ref{eq:dbdt}) and 
(\ref{eq:InitialConditionOfab}), 
is our first result.

We would like to make two comments:
First, the parameter setting 
$\{S_h = 1/T_h,S_c=S_u=0\}$ 
gives a kind of an integral fluctuation relation 
\cite{Seifert2012,Holubec2021}:
\be
\left\langle \exp\left(\frac{Q_h}{T_h}\right) 
\right\rangle 
= \exp\left(
\frac{1}{\gamma} \int_{0}^{t_1} \lambda(t')dt'
\right)
\,.
\ee
Secondly, when we consider the case $T_h = T_c = T$ and  set
$\{ S_h = S_c = S_w , S_u = - S_w \}$, 
$M$ reduces to a moment generating function of $W$,   
$\langle e^{S_w W} \rangle$, which can be shown to 
be equivalent to the result obtained in Ref.~\cite{Speck2011}.

\section{\label{sec:2nd}Variances and Covariances}

We can calculate averages, variances and covariances of 
$Q_h$, $Q_c$ and $\Delta U$ by 
differentiating $\log M$ with respect to ${\bm S} = \{S_h,S_c,S_u\}$
and then setting ${\bm S} = {\bm 0}$.

We first calculate averages and check whether they agree with 
previously obtained results \cite{Schmiedl2008};
\ba
\left\langle Q_h\right\rangle
&=& \left.\frac{\partial\log M}{\partial S_h}\right\vert_{\bm S={\bm 0}}
=
\frac{ T_h}{\gamma} \int_{0}^{t_1} \lambda(t')  dt'
- 
\frac{1}{\gamma} \int_{0}^{t_1} \lambda^2(t') w(t')dt'
\nonumber\\
&=& \frac{ T_h}{2} \log\frac{w_b}{w_a}
- \frac{\gamma}{4}\int_{0}^{t_1} \frac{\dot{w}^2(t')}{w(t')} dt'
\,.
\label{eq:avQh}
\\
\langle Q_c\rangle 
&=& \left.\frac{\partial\log M}{\partial S_c}
\right\vert_{{\bm S}={\bm 0}}
= \frac{ T_c}{2} \log\frac{w_a}{w_b}
- \frac{\gamma}{4}\int_{t_1}^{t_1+t_3} \frac{\dot{w}^2(t')}{w(t')} dt'
\,,
\label{eq:avQc}
\\
\langle W\rangle &=& 
\langle Q_h + Q_c - \Delta U\rangle = \langle Q_h + Q_c \rangle
\nonumber\\
&=&   
\frac{ (T_h - T_c)}{2} \log\frac{w_b}{w_a}
- \frac{\gamma}{4}\int_{0}^{t_1+t_3} \frac{\dot{w}^2(t')}{w(t')} dt'
\,.
\label{eq:avW}
\ea
In the 2nd line of Eq.~(\ref{eq:avQh}), by using Eq.~(\ref{eq:dwdt}), 
we express the integrand in terms of $w$ and $\dot{w}$ rather than 
$\lambda$ because the protocol of the system is given by $w(t)$ and 
not by $\lambda(t)$.
The first term in the 2nd line gives the quasi-static limit.

All these results, Eqs.~(\ref{eq:avQh})$\sim$(\ref{eq:avW}), 
agree with those obtained in Ref.~\cite{Schmiedl2008}.
We note that in the quasistatic limit, 
$\frac{\langle W\rangle}{\langle Q_h\rangle}$ becomes 
the Carnot efficiency;
\be
\frac{\langle W\rangle}{\langle Q_h\rangle} 
= 1 - \frac{T_c}{T_h} 
 + \cdots
\,\,.
\ee

As for second moments, we need to differentiate
 $b(t)$ with respect to 
${\bm S} = \{S_h,S_c,S_u\}$:
\be
b_\alpha(t)=  \left.\frac{\partial b(t)}{\partial S_\alpha}
\right\vert_{{\bm S}= {\bm 0}}
\,,\quad \alpha = h, c, u
\,.
\ee
$b_\alpha(t)$ satisfy the following equations: 
\ba
\frac{db_h(t)}{dt} &=& 
 - \frac{2}{\gamma} \lambda(t) b_h(t) + 
\frac{2}{\gamma} \theta(t_1-t) \lambda(t) w(t) 
\Big( 2  T_h - \lambda(t) w(t) \Big)
\,,\quad
b_h(0) = 0
\,,
\label{eq:dbhdt}
\\
\frac{db_c(t)}{dt} &=& 
 - \frac{2}{\gamma} \lambda(t) b_c(t) + 
\frac{2}{\gamma} \theta(t - t_1) \lambda(t) w(t) 
\Big( 2  T_c - \lambda(t) w(t) \Big)
\,,\quad
b_c(0) = 0
\,,
\label{eq:dbcdt}
\\
\frac{db_u(t)}{dt}
&=& - \frac{2}{\gamma} \lambda(t) b_u(t)
\,,\quad
b_u(0) = - \lambda(+0) w_a^2
\,.
\label{eq:dbudt}
\ea
The formal solutions of the above equations are given as follows:
\ba
b_h(t) &=& \left\{\begin{array}{ccc}
\frac{w(t)}{T_h}
\int_{0}^{t} ds 
\left(T_h^2 - \frac{\gamma^2}{4} \dot{w}^2(s) \right)
\frac{d}{ds}\left(e^{-\frac{2 T_h}{\gamma}\int_{s}^{t}\frac{dt'}{w(t')}}
\right)
& , & 0 < t < t_1
\\
  \frac{b_h(t_1)}{w(t_1)}w(t) 
e^{- \frac{2 T_c}{\gamma} \int_{t_1}^{t} \frac{dt'}{w(t')}}
& , & t_1 < t < t_1 + t_3
\end{array}
\right.
\label{eq:bh}
\\
b_c(t) &=& \left\{\begin{array}{ccc}
0 & , & 0 < t < t_1
\\
\frac{w(t)}{T_c}
\int_{t_1}^{t} ds 
\left( T_c^2 - \frac{\gamma^2}{4}\dot{w}^2(s)\right)
\frac{d}{ds}\left(e^{-\frac{2 T_c}{\gamma}\int_{s}^{t}
\frac{dt'}{w(t')}}\right)
& , & t_1 < t < t_1 + t_3
\end{array}
\right.
\label{eq:bc}
\\
b_u(t) &=& - \left(T_h - \frac{\gamma}{2}\dot{w}(+0) \right) w(t)
e^{-\frac{2}{\gamma}\int_{0}^{t} dt' \frac{T(t')}{w(t')}}
\,,
\label{eq:bu}
\ea
which we use when considering the quasi-static limit.

Using these functions, 
we get the following expressions for 2nd moments;
\ba
{\rm Var}(Q_h)
&=& \left.\frac{\partial^2\log {M}}{\partial S_h^2}
\right\vert_{{\bm S}= {\bm 0}}
=
\frac{2 T_h}{\gamma} \int_{0}^{t_1} \lambda^2(t') w(t') dt'
- \frac{2}{\gamma} \int_{0}^{t_1} \lambda^2(t') b_h(t') dt'
\nonumber\\
&=&
\frac{2}{\gamma}\int_{0}^{t_1}
\frac{1}{w^2(t')}
\left( T_h - \frac{\gamma}{2} \dot{w}(t') \right)^2
\Big(  T_h w(t') - b_h(t') \Big) dt'
\nonumber\\
&=&
( T_h)^2 + \cdots
\,.
\label{eq:dQhdQh}
\ea
\ba
{\rm Var}(Q_c)
&=& \left.\frac{\partial^2\log {M}}{\partial S_c^2}
\right\vert_{{\bm S}={\bm 0}}
\nonumber\\
&=&
\frac{2}{\gamma}\int_{t_1}^{t_1+t_3}
\frac{1}{w^2(t')}
\left(T_c - \frac{\gamma}{2} \dot{w}(t') \right)^2
\Big(  T_c w(t') - b_c(t') \Big) dt'
\nonumber\\
&=& ( T_c)^2 + \cdots
\,.
\label{eq:dQcdQc}
\ea
\ba
{\rm Var}(\Delta U)
&=& \left.\frac{\partial^2\log {M}}{\partial S_u^2}
\right\vert_{{\bm S}={\bm 0}}
= \left(\lambda(+0) w_a \right)^2+b_u(t_1+t_3+0)\lambda(+0)
\nonumber\\
&=& \left( T_h - \frac{\gamma}{2}\dot{w}(+0) \right)^2
\left( 1 - e^{-\frac{2}{\gamma}\int_{0}^{t_2+t_3} dt' \frac{T(t')}{w(t')}} 
\right)
= ( T_h)^2 + \cdots
\,.
\label{eq:dUdU}
\ea
\ba
{\rm Cov}(Q_h,Q_c) &=& 
\left.\frac{\partial^2\log\hat{M}}
{\partial S_h \partial S_c}\right\vert_{{\bm S}={\bm 0}}
= - \frac{1}{\gamma}\int_{t_1}^{t_1+t_3}\lambda^2(t') b_h(t') dt'
\nonumber\\
&=& 
- \frac{1}{\gamma}  \frac{b_h(t_1)}{w(t_1)}
\int_{t_1}^{t_1+t_3} dt' \frac{(T_c - \gamma \dot{w}(t')/2)^2}{w(t')}
e^{- \frac{2 T_c}{\gamma} \int_{t_1}^{t'} \frac{ds}{w(s)}}
\nonumber\\
&=&  
-\frac{ T_h  T_c}{2} + \cdots
\,.
\label{eq:dQhdQc}
\ea
\ba
{\rm Cov}(\Delta U,Q_h) &=& 
\left.\frac{\partial^2\log {M}}{\partial S_h \partial S_u}
\right\vert_{{\bm S}={\bm 0}}
\nonumber\\
&=&
- \frac{1}{\gamma}\int_0^{t_1}\lambda^2(t') b_u(t') dt'
+\frac{1}{2}\lambda(t_1+t_3+0)b_h(t_1+t_3+0) 
\nonumber\\
&=&
- \frac{1}{2}\left( 1 - \frac {\gamma}{2}\frac{\dot{w}(+0)}{ T_h}  \right) 
\int_0^{t_1} dt' 
\left(  T_h - \frac{\gamma \dot{w}(t')}{2} \right)^2
\frac{d}{dt'}e^{-\frac{2 T_h}{\gamma}\int_0^{t'}\frac{ds}{w(s)}}
\nonumber\\
&&
+ \left(  T_h - \frac{\gamma \dot{w}(+0)}{2} \right)
\frac{b_h(t_1)}{2 w_b}e^{- \frac{ 2 T_c}{\gamma}\int_{t_1}^{t_1+t_3} \frac{ds}{w(s)}}
 = \frac{( T_h)^2}{2} + \cdots
\,.
\label{eq:dQhdU}
\ea
\ba
{\rm Cov}(\Delta U,Q_c) &=&
\left.\frac{\partial^2\log {M}}{\partial S_c \partial  S_u}
\right\vert_{{\bm S}={\bm 0}}
\nonumber\\
&=&
- \frac{1}{\gamma}\int_{t_1}^{t_1+t_3}\lambda^2(t') b_u(t') dt'
+ \frac{1}{2}\lambda(t_1+t_3+0)b_c(t_1+t_3+0) 
\nonumber\\
&=&
- \frac{T_h - \frac{\gamma}{2}\dot{w}(+0)}{2 T_c}
 \int_{t_1}^{t_1+t_3} dt' 
\left( T_c - \frac{\gamma}{2} \dot{w}^2(t')\right)^2 
\frac{d}{dt'} 
\left(e^{- \frac{2}{\gamma} \int_{0}^{t'}\frac{T(s)}{w(s)}ds }\right)
\nonumber\\
&&
+ \frac{T_h - \frac{\gamma}{2} \dot{w}(+0)}{2 T_c}
\int_{t_1}^{t_1+t_3} ds 
\left( (T_c)^2 - \frac{\gamma^2}{4}\dot{w}^2(s) \right)
\frac{d}{ds} 
\left(e^{- \frac{2 T_c}{\gamma} \int_{s}^{t_1+t_3}
\frac{dt'}{w(t')}}\right)
\nonumber\\
&&= 
\frac{ T_h  T_c}{2} + \cdots
\,.
\label{eq:dQcdU}
\ea
Eqs.~(\ref{eq:dQhdQh})$\sim$(\ref{eq:dQcdU}) are our second results.
The term in the rightmost side of each equation shows the quasi-static limit 
where terms of the order of
$O(\gamma)$,
$O\left(\exp\left(-\frac{2 T_h}{\gamma}\int_0^{t_1}\frac{ds}{w(s)}\right)
\right)$,
$O\left(\exp\left(-\frac{2 T_c}{\gamma}\int_{t_1}^{t_1+t_3}
\frac{ds}{w(s)}\right)
\right)$
are neglected.
We note that the quasi-static limit depends only on $T_h$ and $T_c$.

The variance and covariance of $W$ are obtained from the above results with 
the use of Eq.~(\ref{eq:1st}):
\ba
{\rm Var}(W)&=&{\rm Var}(Q_h)+{\rm Var}(Q_c)+{\rm Var}(\Delta U)
\nonumber\\
&&\quad 
+2\Big({\rm Cov}(Q_h,Q_c)-{\rm Cov}(\Delta U,Q_c)-{\rm Cov}(\Delta U,Q_h) \Big)
\nonumber\\
&=& \left( T_h -  T_c \right)^2 + \cdots
\,.
\label{eq:dWdW}
\ea
\ba
{\rm Cov}(Q_h,W) &=&
 {\rm Var}(Q_h) + {\rm Cov}(Q_h,Q_c) -  {\rm Cov}(\Delta U,Q_h)
\nonumber\\
&=&
\frac{ T_h (  T_h -  T_c )}{2} + \cdots
\,.
\label{eq:dQhdW}
\ea
\ba
{\rm Cov}(Q_c,W) &=&
 {\rm Var}(Q_c) + {\rm Cov}(Q_h,Q_c) -  {\rm Cov}(\Delta U,Q_c)
\nonumber\\
&=&
  - T_c (  T_h -  T_c ) + \cdots
\,.
\label{eq:dQcdW}
\ea
We here write down only the expressions in the quasi-static limit. 
We note that the quasi-static limit of ${\rm Var}(W)$ is 
consistent with the previously obtained result 
(see Eq.~(S12) in the supplemental material of Ref.~\cite{Holubec2018}).

\section{\label{sec:discussion} Discussion}
Figs. 1 $\sim$ 3,
 show how ${\rm Var}(W)$, ${\rm Var}(Q_h)$, and 
${\rm Cov}(Q_h,W)$ (denoted by solid lines ) approach  
their quasi-static limit
 (denoted by horizontal dashed lines)
, respectively.
The horizontal axis shows the time interval of a 
cycle of the heat engine.
The following values are used:
\be
 \gamma = 1\,,\,
 w_a = 0.5\,,\, w_b = 1\,,\,
 T_c = 1\,,\, T_h = 2
\,.
\ee
The dots show results obtained from  
the ensemble of 10000 
trajectories 
which are calculated with the use of Eq.~(\ref{eq:SDE})
with $n = 1001$.
The agreement between solid lines and dots 
demonstrates the correctness of the procedure in Sect.~\ref{sec:logM}.

\begin{figure}[!h]
\centering\includegraphics[width=3.0in]{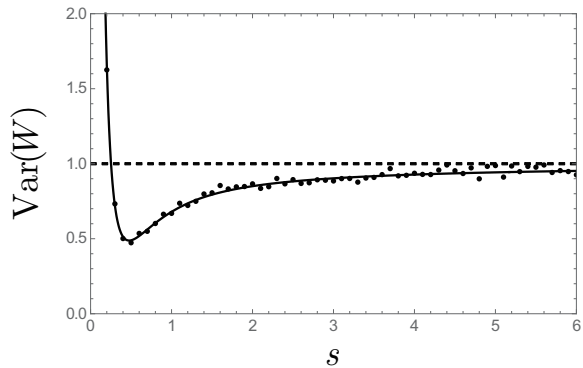}
\vspace*{8pt}
\caption{
${\rm Var}(W)$ plotted against $s$ defined in (\ref{eq:s})
is denoted by the solid line.
The horizontal dashed line shows the quasi-static limit.
Dots are calculated from the ensemble of 10000 
trajectories.
}
\label{fig:dWdW}
\end{figure}

\begin{figure}[!h]
\centering\includegraphics[width=3.0in]{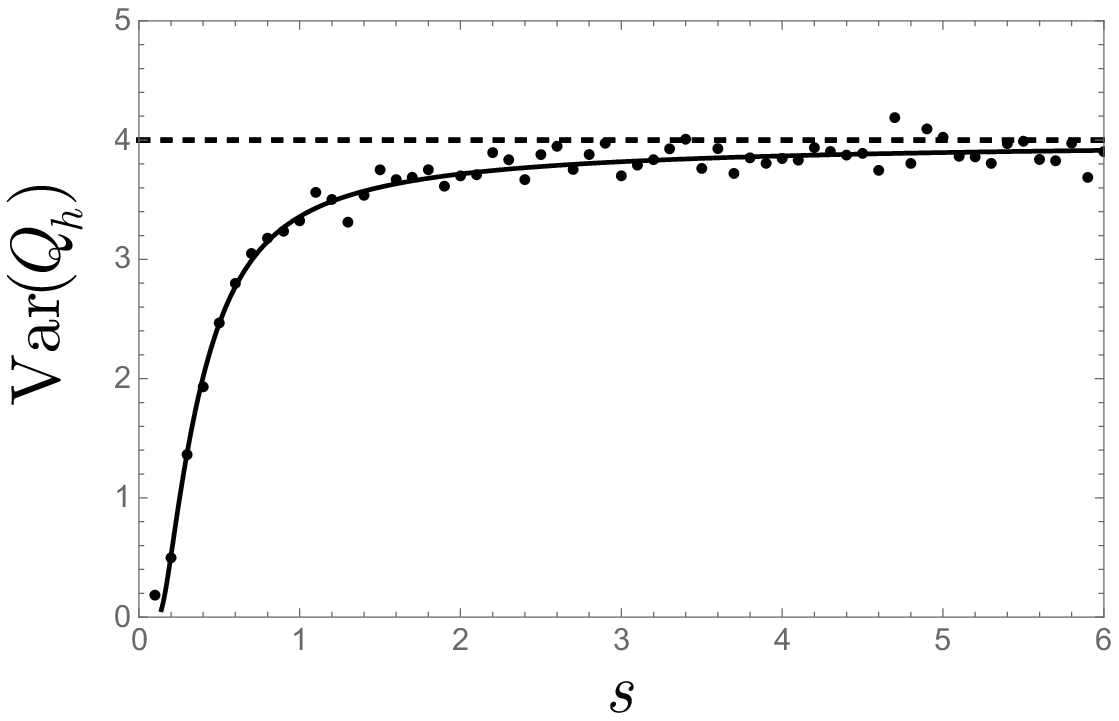}
\vspace*{8pt}
\caption{
${\rm Var}(Q_h)$ plotted against $s$ defined in (\ref{eq:s})
is denoted by the solid line.
The horizontal dashed line shows the quasi-static limit.
Dots are calculated from the ensemble of 10000 trajectories.
}
\label{fig:dQhdQh}
\end{figure}

\begin{figure}[!h]
\centering\includegraphics[width=3.0in]{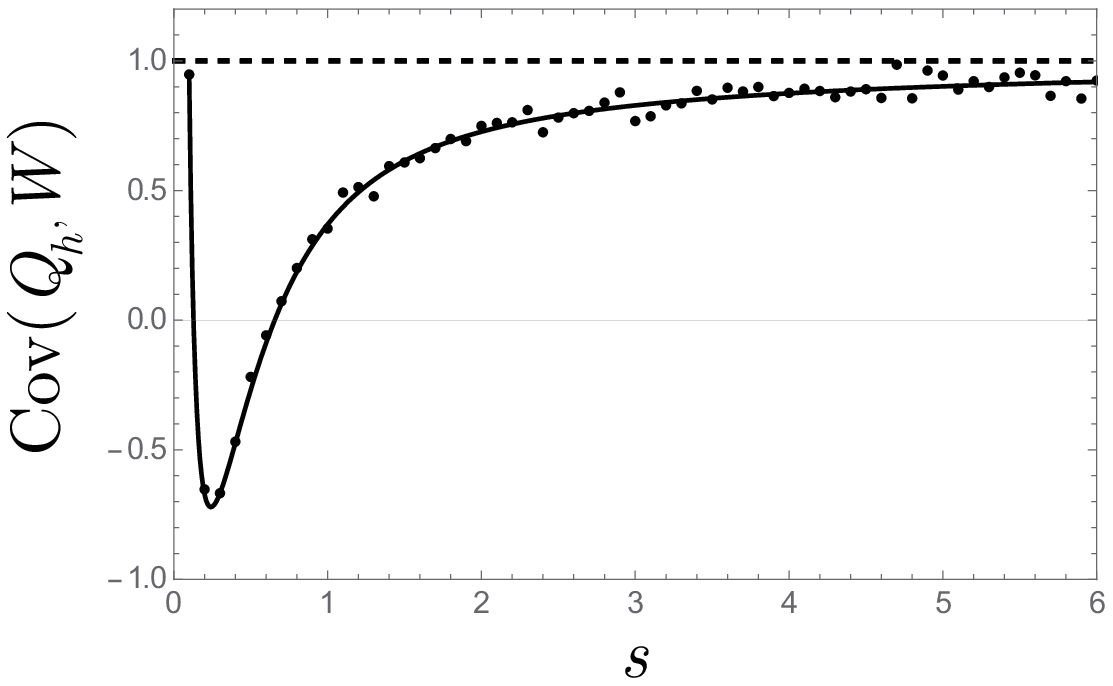}
\vspace*{8pt}
\caption{
${\rm Cov}(Q_h,W)$ plotted against $s$ defined in (\ref{eq:s})
is denoted by the solid line.
The horizontal dashed line shows the quasi-static limit.
Dots are calculated from the ensemble of 10000 trajectories.
}
\label{fig:dWdQh}
\end{figure}

\begin{figure}[!h]
\centering\includegraphics[width=3.0in]{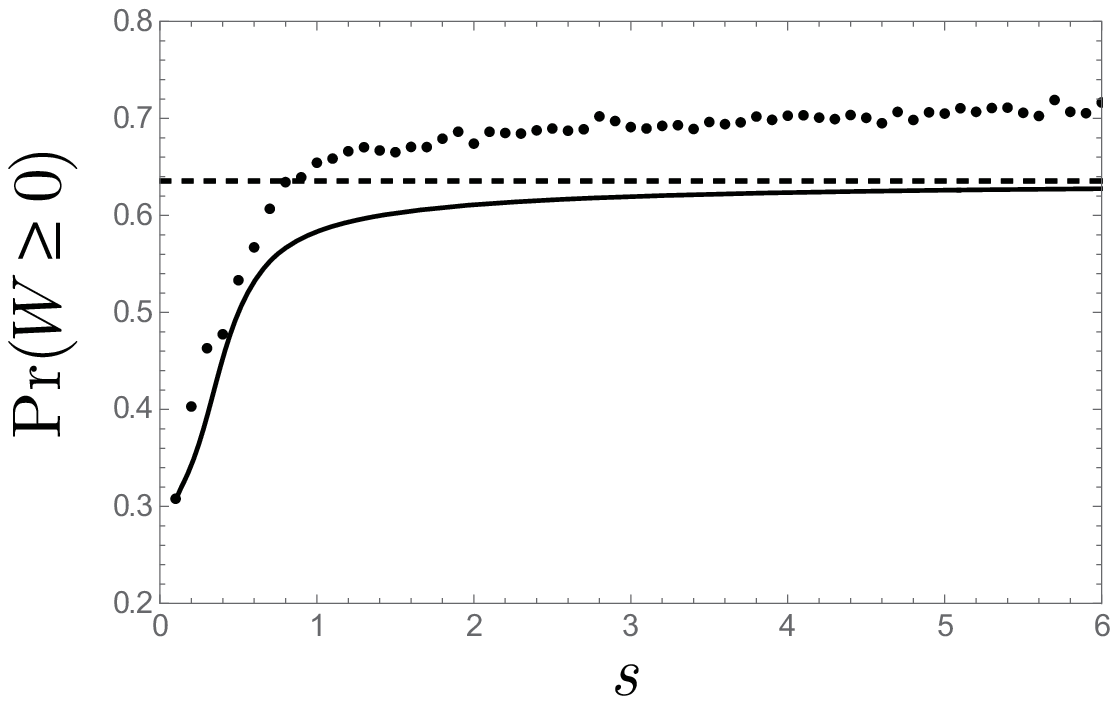}
\vspace*{8pt}
\caption{
The approximate value of 
${\rm Pr}\left(W \geq 0 \right)$
plotted against $s$ defined in (\ref{eq:s})
is denoted by the solid line.
The horizontal dashed line shows the quasi-static limit.
Dots are calculated from the ensemble of 10000 trajectories.
}
\label{fig:PrbW}
\end{figure}

What can we say about the performance of 
a stochastic heat engine
from the results obtained in this paper?
Ref.~\cite{Rana2014} stated that
the full probability distribution is necessary 
for the proper analysis of the system
because fluctuations dominate the mean values.
Lacking the full probability distribution, 
we have sought possible candidates 
which can be calculated using 2nd moments.
We suggest the approximate probability for $W$ to 
exceeds a given threshold $W_0$:  
\ba
 {\rm Pr}\left(W \geq W_0 \right) &=& 
{\rm Pr}\left( \frac{W- \langle W\rangle}{\sqrt{{\rm Var}(W)}} \geq 
-\frac{1}{k} + \frac{W_0}{\sqrt{{\rm Var}(W)}} \right)
\nonumber\\
&\approx& 
\frac{1}{\sqrt{2\pi}}\int_{-\frac{1}{k} + \frac{W_0}{\sqrt{{\rm Var}(W)}}}^{\infty} 
e^{-\frac{z^2}{2}} dz
\label{eq:prbW}
\ea
where 
\be
 k = \frac{\sqrt{{\rm Var}(W)}}{\langle W\rangle} = 
 \frac{2}{\log\left(\frac{w_b}{w_a}\right)} + \cdots 
\ee
is the coefficient of variation of $W$.
In the rightmost side of Eq.~(\ref{eq:prbW}),
we approximate the distribution of $W$ by Gaussian.
We know it is not Gaussian but hope 
the details (e.g. skewness or long tails  etc.)
may not significantly affect the integrated value.
( As for the distribution of $W$, see 
Fig.22 in Ref.~\cite{Rana2014} though the protocol of 
$\lambda(t)$ is different from that used here.) 

Fig.\ref{fig:PrbW} shows the results for $W_0 = 0$.
Solid line denotes the rightmost side of Eq.~(\ref{eq:prbW}).
Dots show the ratio of trajectories with $W \ge 0$ 
obtained from the ensemble.
We can see that the approximation 
reproduces the overall tendency
though the agreement is moderate and
further investigation is certainly necessary.

\section{\label{sec:summary}Summary}

In this paper, we consider a simple model of a stochastic 
heat engine, which consists of a single Brownian particle moving 
in a one-dimensional periodically breathing  harmonic potential.
Overdamped limit is assumed (see Eq.~(\ref{eq:EOM-2})).
Expressions of second moments (variances and covariances ) of heat and work 
are obtained in the form of integrals 
(see Eqs.~(\ref{eq:dQhdQh})$\sim$(\ref{eq:dQcdW})), 
whose integrands contain 
functions satisfying certain differential equations 
(see Eqs.~(\ref{eq:dbhdt})$\sim$(\ref{eq:dbudt})).
The results in the quasi-static limit are simple functions of 
temperatures of hot and cold thermal baths.
The coefficient of variation of the work is suggested to 
give an approximate probability for the work to 
exceeds a given threshold, which may serve
as an index of the performance of a stochastic heat engine. 
In the course of derivation, we get the expression of the cumulant-generating 
function (see Eq.~(\ref{eq:logM-3})). 

Comments on further study are in order;
Including a linear term $c(t) x$ in $U(x,t)$ could be addressed
within the present formulation.
Considering the underdamped case seems also interesting.
To consider $U(x,t)$ other than a harmonic potential 
would require a different formulation.
It may be worth investigating to what extent do quasi-static limits 
obtained here hold.

\section*{Acknowledgements}

S.I. thanks N. Ikeda for enlightening him on a stochastic differential 
equation.
S.I. is supported by a Ryukoku University Research Fellowship (2021).


%

\vspace{0.2cm}
\noindent


\let\doi\relax


\end{document}